\documentclass[sigconf, nonacm]{acmart}

\AtBeginDocument{%
  \providecommand\BibTeX{{\normalfont B\kern-0.5em{\scshape i\kern-0.25em b}\kern-0.8em\TeX}}}

\graphicspath{{./figures/}}

\usepackage{multirow}
\usepackage{booktabs}
\usepackage{pifont}   
\usepackage{dblfloatfix}  

\begin{document}

\title{Lost in Aggregation: A Multi-Scale Diagnostic Benchmark for LLM Spatial Navigation}

\author{Yuhan Jiang}
\email{yuhan.jiang@tum.de}
\affiliation{%
  \institution{Technical University of Munich}
  \city{Munich}
  \country{Germany}}

\author{Peng Luo}
\authornote{Corresponding author.} 

\email{pengluo@mit.edu}
\affiliation{%
  \institution{Massachusetts Institute of Technology}
  \city{Cambridge}
  \country{USA}}

\author{Liqiu Meng}
\email{liqiu.meng@tum.de}
\affiliation{%
  \institution{Technical University of Munich}
  \city{Munich}
  \country{Germany}}

\renewcommand{\shortauthors}{Jiang et al.}

\begin{abstract}
Large language models (LLMs) are increasingly deployed as planners and
assistants in tasks with inherent spatial structure, such as navigation and
route planning, yet they remain brittle in sequential spatial reasoning.
We ask not merely \emph{whether}
LLMs fail at navigation but \emph{where} in the spatial-cognition pipeline
they get lost. We introduce a multi-scale diagnostic benchmark that
decomposes maze navigation into three cognitive levels drawn from human
spatial cognition: \textbf{Fine} (local passability), \textbf{Meso}
(junction topology), and \textbf{Macro} (global goal direction). We
evaluate three instruction-tuned chat LLMs (GPT-4o, DeepSeek-V3,
Llama-3.3-70B) on
1{,}050 topology-annotated mazes spanning seven sizes ($3{\times}3$ to
$30{\times}30$) and three difficulty tiers. The benchmark is organized as
three modules. \emph{(i)~Input acquisition:} among four input formats,
structured coordinate text is the most navigable, far surpassing
rendered images. \emph{(ii)~Multi-scale
representation:} end-to-end one-shot navigation collapses to near zero by
$10{\times}10$ for every model, yet the same models respond to isolated
single-level probes (Fine, Meso, Macro) at $30$--$75\%$ far beyond that
size. A multi-hot first-error analysis localizes failures to Meso junction
choices ($59\%$) and Fine perception ($39\%$), with global direction almost
never at fault ($1\%$). The barrier
is therefore the cross-scale \emph{aggregation} of individually available
competences over a long sequential plan, not any single perceptual deficit.
\emph{(iii)~Hierarchical route planning:} delegating per-step execution to
a deterministic walker and querying the LLM only at junctions, with an
explicit cell-type prompt, lifts GPT-4o success by up to $92$ points at mid
sizes, but the same scaling wall re-emerges by $30{\times}30$. We release
the benchmark, mazes, and code as a reusable diagnostic instrument for
spatial reasoning in LLMs, available at
\url{https://yuhanjiang415.github.io/lost-in-aggregation/}.
\end{abstract}



\maketitle

\section{Introduction}
\label{sec:intro}

Large language models (LLMs) are increasingly used as planners and
assistants in tasks with inherent spatial structure, such as turn-by-turn
navigation, route planning, and map-based decision making. They also serve
as the reasoning core of geospatial agents that query maps and spatial databases
\cite{Li2025MapQA, Yu2025SpatialRAG, Wang2026Disaster}. Yet a growing body of evidence shows
that even strong models are brittle in spatial reasoning, particularly in
\emph{sequential} settings that require maintaining a coherent spatial
state across many steps \cite{Shiri2024, Martorell2025, Ramakrishnan2025SpatialCog}. Knowing
\emph{that} LLMs fail at navigation is, by now, unsurprising. The open and
more actionable question is \emph{where} in the spatial-cognition pipeline
they get lost: Are the deficits due to an inability to perceive local geometry, read topological structure at branch points, hold a global heading, or \emph{integrate} these across sizes over a long plan?
An aggregate
success rate cannot distinguish these hypotheses, and they imply opposite
fixes: better perception, better topological reasoning, or a different
division of labor between the model and a surrounding system.

\begin{figure}[t]
    \centering
    \includegraphics[width=\linewidth]{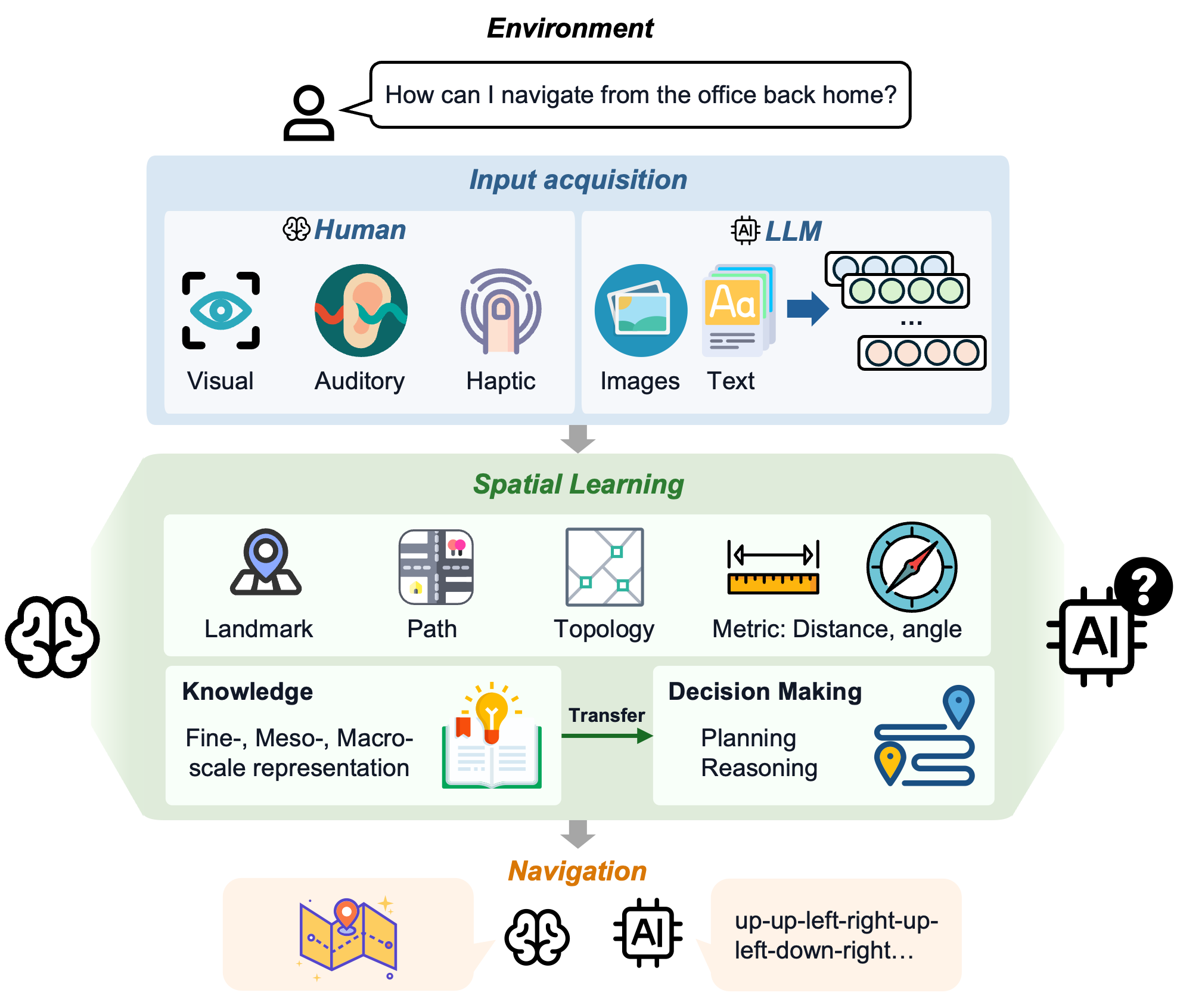}
    \caption{\textbf{From spatial input to navigation: where LLMs differ
    from human spatial cognition.} Starting from a natural-language
    navigation request, the pipeline traces \emph{input acquisition} (humans
    use vision, audition, and haptics; an LLM receives only image and text
    tokens), \emph{spatial learning} (organizing the signal into Landmark /
    Path / Topology / Metric primitives, which we cast as a three-level
    Fine / Meso / Macro representation), and \emph{decision making} that
    yields an executable move sequence. The benchmark diagnoses LLM
    navigation by decomposing it into the same representational levels that
    structure human spatial cognition.}
    \Description{A conceptual flowchart comparing the human and LLM spatial
    cognition pipelines, from input acquisition through spatial learning
    (Fine, Meso, Macro levels) to decision making and an executable move
    sequence.}
    \label{fig:framework}
\end{figure}

We approach this question through the lens of human spatial cognition,
long described as a multi-scale phenomenon. People reason about space
differently at the scale of immediately perceivable surroundings and at the
scale of environments assembled from many observations
\cite{Montello1993, Freundschuh1997, Wolbers2010, Peer2025CognitiveMaps}. We
operationalize this into three cognitive levels that a navigator must
sustain simultaneously (Figure~\ref{fig:framework}):
\textbf{Fine} (local positioning, which adjacent cells are passable),
\textbf{Meso} (topological mapping, which branch to take at a junction and
when a corridor is a dead-end), and \textbf{Macro} (goal orientation, the
global heading toward the destination). Maze navigation is a natural
testbed: it requires all three levels at once, it scales continuously in
size, and every intermediate decision has a verifiable ground truth.

We instantiate this view as a multi-scale diagnostic benchmark with three
\emph{modules} that mirror the stages of the navigation pipeline:
(1)~\textbf{input acquisition}: which textual or visual encoding of space
an LLM reasons over best; (2)~\textbf{multi-scale representation}: a
scale-resolved decomposition of where navigation breaks, including each
cognitive level probed \emph{in isolation}; and (3)~\textbf{hierarchical
route planning}: how much low-level control must be delegated from the LLM
to a deterministic algorithm before navigation is restored. These three
modules answer three research questions:

\begin{itemize}
  \item \textbf{RQ1.} Which spatial input format best supports LLM navigation?
  \item \textbf{RQ2.} Where, across spatial sizes and across the Fine /
  Meso / Macro levels, do LLMs fail?
  \item \textbf{RQ3.} Which spatial control should a hybrid navigation agent
  delegate to the LLM, and which to the algorithm?
\end{itemize}

\paragraph{Contributions.} (1)~A cognitively grounded, multi-scale
diagnostic benchmark for LLM spatial navigation: $1{,}050$
topology-annotated mazes across seven sizes and three difficulty tiers,
four input formats, isolated single-level probes with non-spatial
controls, and a junction-delegation harness, released as code and data.
(2)~Evidence that scaling failure is fundamentally a failure of
\emph{cross-scale aggregation}, not of local perception or global
direction: isolated single-level competences survive far beyond the size
at which end-to-end navigation collapses, the coupled Meso$\times$Macro probe
decays faster than its Meso component (and comparably to Macro), and a multi-hot
first-error taxonomy places the
blame on Meso and Fine, almost never on Macro. (3)~A delegation study
quantifying how much low-level control, and how much explicit
topological framing, must move from the LLM to the algorithm before
navigation recovers, with implications for hybrid GIS and wayfinding agents.

\section{Related Work}
\label{sec:related}

\paragraph{Spatial reasoning in LLMs.}
Empirical analyses report that large (multimodal) models are unreliable on
tasks requiring spatial relations, orientation, and multi-step spatial
inference, often performing at near-chance levels once several relations must be
composed \cite{Shiri2024}. Textual spatial-reasoning benchmarks make the
multi-hop case explicit: accuracy degrades steeply as the number of composed
relations grows \cite{Shi2022StepGame, Mirzaee2021SpartQA}. Recent work
probes the internal spatial representations LLMs form during grid-world
navigation and finds them partial and unstable across steps
\cite{Martorell2025}. Models likewise struggle to assemble a global map from
purely local descriptions \cite{Xia2025MapWorld}. Related efforts propose
text-based map encodings to ground spatial reasoning and navigation
\cite{Zhang2024}. Beyond evaluation, recent methods aim to strengthen spatial
reasoning through depth-grounded region representations in vision-language
models \cite{Cheng2024SpatialRGPT} and visualization-of-thought prompting
\cite{Wu2024VoT}. Surveys and benchmarks cover this area across scales and
modalities \cite{Feng2025SpatialSurvey, Yang2025SpatialCog, Xu2025SpatialBench}.
A complementary concern is whether
apparent competence reflects reasoning or memorized patterns: counterfactual
variants of standard tasks expose large drops, cautioning that isolated
single-question accuracy can overstate genuine capability \cite{Wu2024Counterfactual}.

Most such studies report a single aggregate score. Our work differs in
granularity: rather than asking how well a model navigates, we \emph{decompose}
navigation into Fine / Meso / Macro levels and probe each in isolation, so the
benchmark localizes the deficit. A delegation study then tests which level a
surrounding system must take over.

\paragraph{Maze and grid-world navigation benchmarks.}
Mazes and grid worlds are a standard controlled setting for studying
sequential spatial decision making: they admit exact ground truth at every
step and scale continuously in difficulty. Configurable generators also make
them easy to standardize \cite{Ivanitskiy2023MazeDataset}. The closest
prior benchmark is \textsc{SpatialEval} \cite{Wang2024SpatialEval}, whose
Maze-Nav task evaluates LLMs and multimodal models across text, vision, and
vision--text encodings. Its finding that textual input tends to beat rendered
images directly anticipates our Module~1 result (Section~\ref{sec:res-rq1}),
and we treat that finding as confirmed and quantified rather than novel.
\textsc{MazeEval} likewise probes sequential decision making with
coordinate-based feedback \cite{Einarsson2025MazeEval}. Others instead fine-tune
LLMs to improve maze solving directly, for example with reinforcement learning
\cite{Dao2025AlphaMaze}. Our central concern
is more general than maze success per se: it is \emph{planning} over a long
horizon. The canonical result there is that LLMs degrade sharply as plan
length grows even when single-step reasoning is intact
\cite{Valmeekam2023PlanBench}, the planning analogue of the aggregation
deficit we localize. Prompting strategies that externalize intermediate
reasoning, such as chain-of-thought \cite{Wei2022CoT}, deliberate search over
thoughts \cite{Yao2023ToT}, and language-model planning with a world model
\cite{Hao2023RAP}, reduce but do not remove these failures. A complementary
response is architectural: rather than expect an LLM to plan end-to-end,
recent frameworks pair it with external actions and solvers
\cite{Yao2023ReAct, Kambhampati2024LLMModulo}, which motivates our Module~3
delegation study. Beyond symbolic grids, embodied vision-and-language
navigation studies established the basis for following routes in photorealistic environments
\cite{Anderson2018R2R}, increasingly using LLMs as the reasoning core
\cite{Zhou2024NavGPT}, with LLM agents now tackling object- and city-scale
navigation \cite{Wu2024VoroNav, Mei2025UrbanNav}. That setting's perceptual
demands are different from the symbolic, fully-observed reasoning we isolate.

Prior benchmarks largely score end-to-end task success at a fixed size.
In contrast, we introduce three additional axes: a systematic \emph{size} sweep
($3{\times}3$ to $30{\times}30$) that exposes where competence breaks,
\emph{isolated single-level probes} that separate availability of a
competence from its use inside navigation, and non-spatial controls over the
identical maze text that separate spatial reasoning from generic long-context
reading load \cite{Liu2024LostMiddle}.

\paragraph{Geospatial question answering and retrieval.}
A parallel line equips LLMs with spatial data through question answering
over maps \cite{Li2025MapQA} and spatial retrieval-augmented generation
\cite{Yu2025SpatialRAG}. Whether vision-language models genuinely read map
images, rather than lean on textual priors, remains contested
\cite{Xing2025ReadMaps}. These systems presuppose that the LLM can reason
over the spatial structure they surface. Our diagnosis of \emph{where} that
reasoning breaks down is directly relevant to designing such pipelines, as is
our finding that LLMs are best used as junction-level decision makers inside
an algorithmic skeleton.

\paragraph{Multi-scale spatial cognition.}
Cognitive science describes human spatial knowledge as scale dependent:
figural, vista, environmental, and geographic spaces engage different
processes \cite{Montello1993, Freundschuh1997}. Individual navigational
ability reflects the integration of distinct subsystems for landmark,
route, and survey knowledge \cite{Wolbers2010}. We borrow this multi-scale
framing as an organizing \emph{metaphor} (not a claim of cognitive fidelity;
see Limitations), casting Fine / Meso / Macro as loose analogues of local,
topological, and directional spatial knowledge.

\section{Benchmark Design}
\label{sec:benchmark}

\begin{figure*}[t]
    \centering
    \includegraphics[width=\linewidth]{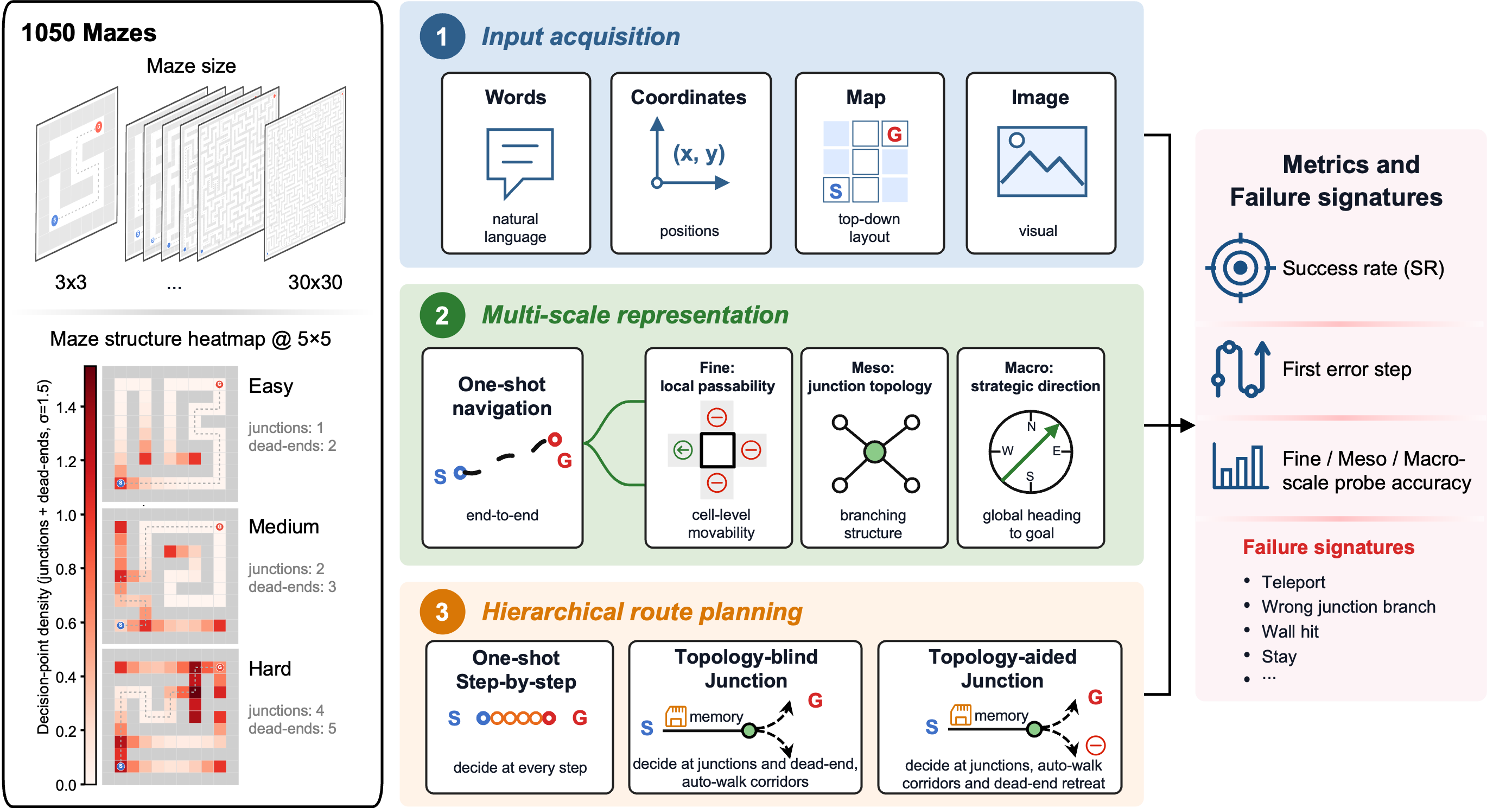}
    \caption{\textbf{Benchmark overview.} \emph{Left:} $1{,}050$ mazes over
    seven effective sizes ($3{\times}3$ to $30{\times}30$) and three
    difficulty tiers, with start (blue) and goal (red) marked.
    \ding{172}~\emph{Input acquisition}: each maze is rendered as Words,
    Coordinates, an ASCII Map, or a rendered Image. \ding{173}~\emph{Multi-scale
    representation}: one-shot navigation is decomposed into \emph{Fine}
    (cell passability), \emph{Meso} (junction topology), and \emph{Macro}
    (global heading), each probed in isolation. \ding{174}~\emph{Hierarchical
    route planning}: One-shot, Topology-blind Junction, and Topology-aided
    Junction delegation regimes. \emph{Right:} metrics (success rate, first
    error step, per-level probe accuracy) and a failure-signature
    vocabulary.}
    \Description{A composite schematic of the benchmark: a maze corpus on
    the left, three numbered modules (input acquisition, multi-scale
    representation, hierarchical planning) in the middle, and a metrics and
    failure-signature panel on the right.}
    \label{fig:overview}
\end{figure*}

Figure~\ref{fig:overview} gives the overall structure: a corpus of mazes
(left) is presented to an LLM through one of four input formats
(module~\ding{172}), navigation is decomposed into three cognitive
levels probed in isolation (module~\ding{173}), and three delegation
regimes contrast how much control the LLM versus an algorithm holds
(module~\ding{174}). Trials are scored by success and by a
failure-signature vocabulary (right).

\subsection{Maze Generation and Annotation}
\label{sec:maze-gen}
Mazes are standard two-dimensional grids ($0=$ path, $1=$ wall) generated
by a mix of depth-first-search (long corridors) and randomized Prim
(many branches) algorithms, with fixed seeds for reproducibility. Including
the outer wall, sizes are $7,11,15,21,31,41,61$, giving \emph{effective}
(open-cell) widths of $3,5,7,10,15,20,30$. Start and goal are placed at
opposite corners. By construction the mazes are trees: \emph{any two cells
are connected by a unique path}, so there are no cycles and every junction
choice is unambiguously right or wrong, sidestepping shortest-path
optimization. For each (size, difficulty) cell we generate $50$ instances,
for $7\times3\times50 = 1{,}050$ mazes in total. Difficulty (easy / medium
/ hard) is controlled by three quantities: the number of decision points on the solution path, dead-end density
(dead-ends per open cell), and a confusion ratio (total decoy-branch length
over solution length). This lets us separate the effect of raw size from
that of local branching complexity. Every maze is pre-annotated with
per-cell passable directions and cell types (corridor, corner, junction,
dead-end), and, along the solution path, the ideal heading and the
goal-reaching branch at each junction: the ground truth used by the
isolated probes and by the module~\ding{174} walker.

\subsection{Three-Dimensional Evaluation}
\label{sec:eval-framework}
Navigation success requires three competences at once.
\textbf{Fine} (positioning) is the ability to read
local passability: which of the four neighbors of the current cell are
open. \textbf{Meso} (topological mapping) is the ability to choose the
goal-reaching branch at a junction and to recognize and retreat from dead
ends. \textbf{Macro} (goal orientation) is the ability to maintain a
correct global heading despite local detours. End-to-end trials are scored
primarily by success rate (SR, reaching the goal via an entirely legal
path) and valid-move ratio (VMR, the fraction of emitted moves that are
grid-legal), with the first-error step recorded for failure analysis
(Table~\ref{tab:metrics}).

\paragraph{Per-level diagnostic metrics and failure signatures.}
Beyond binary success, each emitted trajectory is scored \emph{within} each
level, so that failures can be attributed to a level rather than only counted.
At the Fine level we record the wall-collision rate (moves into a wall) and
teleport rate (non-adjacent jumps), whose complement is a local positioning
accuracy. At the Meso level we record junction accuracy (the fraction of
visited junctions at which a goal-reaching branch is taken), together with
dead-end entry and backtrack-success rates. Since the mazes are trees, a
junction's goal-reaching and shortest-path branches coincide. At the Macro
level we use a windowed progress rate that credits net reduction in goal
distance over each $K$-step window, with $K$ growing with path length.
Averaging over a window keeps necessary local detours from being scored as
heading errors. We report these per-level metrics across sizes in
Section~\ref{sec:res-rq2-perlevel}. Finally, every illegal or sub-optimal move is tagged under a
fixed failure-signature vocabulary grouped by level: at the Fine level a wall
hit, teleport, stay or back-step, or hallucinated start; at the Meso level a
wrong junction branch (scored separately at start, T, and X junctions); and at
the Macro level a wrong global heading. We aggregate these labels in the
first-error analysis (Section~\ref{sec:res-rq2-firsterror}).

\begin{table}[t]
\caption{The three cognitive levels, their isolated single-question
probes, and chance accuracy. A coupled Meso$\times$Macro probe
(junction branch) requires both topological and directional
reasoning in one question.}
\label{tab:dims}
\centering
\small
\begin{tabular}{@{}llc@{}}
\toprule
\textbf{Level} & \textbf{Isolated probe (format)} & \textbf{Chance} \\
\midrule
Fine  & Passable directions (4-choice)        & 0.25 \\
Meso  & Dead-end recognition (yes/no)          & 0.50 \\
Macro & Initial / midcourse heading (4-choice) & 0.25 \\
Meso$\times$Macro & Junction branch (4-choice) & 0.25 \\
\midrule
\multirow{3}{*}{NonSpatial} & Open-cell count (4-choice) & 0.25 \\
                            & Longest wall stretch (4-choice) & 0.25 \\
                            & Row cell listing (free form) & $\approx 0$ \\
\bottomrule
\end{tabular}
\end{table}

\begin{table}[t]
\caption{Evaluation metrics. Overall metrics score the whole trajectory;
per-level metrics attribute failures to Fine, Meso, or Macro.}
\label{tab:metrics}
\centering
\small
\begin{tabular}{@{}llp{0.45\columnwidth}@{}}
\toprule
\textbf{Level} & \textbf{Metric} & \textbf{Definition} \\
\midrule
\multirow{3}{*}{Overall}
 & SR  & reaches the goal via an entirely legal path \\
 & VMR & fraction of emitted moves that are grid-legal \\
 & FES & normalized step index of the first error \\
\midrule
\multirow{2}{*}{Fine}
 & WCR\,/\,TR & rate of wall-collision / teleport (non-adjacent) moves \\
 & PA        & positioning accuracy, $1-\mathrm{WCR}-\mathrm{TR}$ \\
\midrule
\multirow{2}{*}{Meso}
 & JA        & junction accuracy: visited junctions taking a goal-reaching branch \\
 & DER\,/\,BSR & dead-end entry rate / backtrack-success rate \\
\midrule
\multirow{2}{*}{Macro}
 & MPR & progress rate: net goal-distance reduction per $K$-step window \\
 & DDR & direction-drift rate: windows moving away from the goal \\
\bottomrule
\end{tabular}
\end{table}

\subsection{Isolated Probes and Non-Spatial Controls}
\label{sec:isolated-tests}
A central design choice is to test each level \emph{outside} the
navigation loop (Table~\ref{tab:dims}). For a maze of any size we ask single multiple-choice
questions that exercise exactly one level (e.g., ``Which directions can
you move from $(r,c)$?'' for Fine; ``At junction $(r,c)$, which branch is on
the path to the goal?'' for the coupled Meso$\times$Macro probe). Because
the maze encoding (and therefore the input length) is identical
across probes, any difference between probes is attributable to the
\emph{question}, not to context length. We add a \textbf{NonSpatial}
control band of reading-comprehension questions over the same maze text
(count open cells, longest wall run, list a row's open cells) to separate
genuine spatial-cognition decay from generic long-context degradation. The
gap between an LLM's isolated single-level accuracy and its accuracy
\emph{inside} full navigation is itself a measurement of the cost of
cross-scale aggregation (Section~\ref{sec:res-rq2}).

\section{Experimental Setup}
\label{sec:setup}

\paragraph{Models and prompting.}
We evaluate three instruction-tuned chat LLMs through a unified LiteLLM interface:
GPT-4o \cite{OpenAI2024GPT4o}, DeepSeek-V3 \cite{DeepSeek2024}, and
Llama-3.3-70B-Instruct \cite{Llama32024}. All calls use temperature $0$ and
chain-of-thought (CoT) prompting. The one-shot navigation prompt asks the
model to emit the full path as a coordinate sequence
$[(r_1,c_1),(r_2,c_2),\dots]$. We parse it by stripping any
\texttt{<think>} block and extracting the last bracketed coordinate list,
falling back to all coordinate pairs in the cleaned text. We use $n=50$
mazes per (size, difficulty) cell throughout. The one-shot token
budget increases with maze size (from $2{,}048$ tokens at $3{\times}3$ to
$16{,}384$ at $30{\times}30$) so that long paths are never truncated. The
full system prompt and CoT scaffold, and the verbatim text of every probe
and delegation prompt, are reproduced in Appendix
Figure~\ref{fig:supp-framework}.

\paragraph{Module 1: Input acquisition (RQ1).}
Four input formats are compared on small mazes (effective sizes $3,5,7$,
where navigation is not yet saturated by size): natural-language
description (\emph{Words}), an open-cell coordinate list (\emph{Coordinate}),
a top-down ASCII grid (\emph{Map}), and a rendered PNG (\emph{Picture},
sent as a multimodal message). The winning format is fixed for all later
modules. Reported in Section~\ref{sec:res-rq1}.

\paragraph{Module 2: Multi-scale representation (RQ2).}
Using the chosen format, we scale up to all seven sizes and (i)~measure
one-shot end-to-end SR, (ii)~measure isolated single-level probe
accuracy (Fine, Meso, Macro, coupled Meso$\times$Macro, plus NonSpatial
controls), and (iii)~tag the first error of every failed trial under a
multi-hot Fine / Meso / Macro schema. The isolated-probe sweep comprises
$49{,}372$ scored questions, every one of which the single-letter answer
parser resolves successfully (zero parse failures), so the reported
accuracies carry no formatting noise. Reported in
Section~\ref{sec:res-rq2}.

\paragraph{Module 3: Hierarchical route planning (RQ3).}
We contrast three delegation regimes on medium-difficulty mazes at sizes
$7,10,15,20,30$ (GPT-4o and DeepSeek-V3). \emph{One-shot} is the
single-call baseline (the LLM owns Fine + Meso + Macro). In both
\emph{junction} regimes a deterministic walker mechanically traverses
corridors and the LLM is consulted at decision points, with the same
\texttt{NAVIGATION HISTORY} block (past junction choices and discovered
dead-end branches) appended to every prompt. The regimes are matched on
call-site geometry and history budget. They differ only in framing:
\emph{Topology-blind Junction} uses a generic ``navigate the maze'' prompt
with no cell-type hint and no algorithmic dead-end retreat, whereas
\emph{Topology-aided Junction} explicitly tells the model it is at a
junction and the algorithm physically retraces from dead ends.

Each trial is allotted a junction-decision budget proportional to the number
of junctions. Because the blind regime also consults the LLM at dead-ends and
after wall hits, it gets double the per-junction multiplier, matching the two
regimes on their budget \emph{ceiling}. This matches the cap, not the
\emph{realized} call counts: the aided walker survives longer and issues more
calls at large sizes. The regimes are therefore cleanly comparable only at the
smallest size, which we treat as the decisive cell
(Section~\ref{sec:res-rq3}). Reading the SR gap between the regimes reveals
which spatial role the LLM cannot sustain at size.

\subsection{Design Rationale and Threats to Validity}
\label{sec:rationale}
Several choices are deliberate. We use \emph{tree} mazes so that every
junction has exactly one goal-reaching branch, giving each Meso decision an
unambiguous ground truth and removing shortest-path optimization as a
confound. The cost is that we do not yet probe cyclic reasoning, which the
annotation supports as a future extension. The seven \emph{sizes} form a
geometric sweep chosen to bracket the breakpoint of every model rather than
to certify any single size. We hold \emph{prompting} fixed (zero
temperature, a single CoT scaffold) so that size, input format, and
delegation are the only manipulated variables. We deliberately do not tune
prompts per model, which would trade diagnostic clarity for headline scores.
The three \emph{models} span a strong proprietary model (GPT-4o), a strong
open mixture-of-experts model (DeepSeek-V3), and a mid-size open dense model
(Llama-3.3-70B), all queried through one interface.

The design forecloses the most common alternative explanations for
size-driven failure. \emph{Input length} is ruled out by the isolated
probes, which hold the maze text (and thus the token count) fixed while
varying only the question. \emph{Unreadability} of the large-maze encoding
is ruled out by the NonSpatial row-listing control, solved at $92$--$100\%$
at every size. Any residual reading-load effect is addressed by resting
the central claim on the \emph{differential} decay of the coupled spatial
probe relative to its single-level components, not on an absolute
spatial-versus-nonspatial gap. The principal residual threats are those
collected in the limitations: tree-only topology, English-only and
chat-style models, heuristic multi-hot first-error tagging, and exploratory
($n<50$) cells at the largest delegation sizes.

\section{Results and Analysis}
\label{sec:results}

\subsection{Module 1: Choosing an Input Format (RQ1)}
\label{sec:res-rq1}

\begin{figure*}[t]
    \centering
    \includegraphics[width=\linewidth]{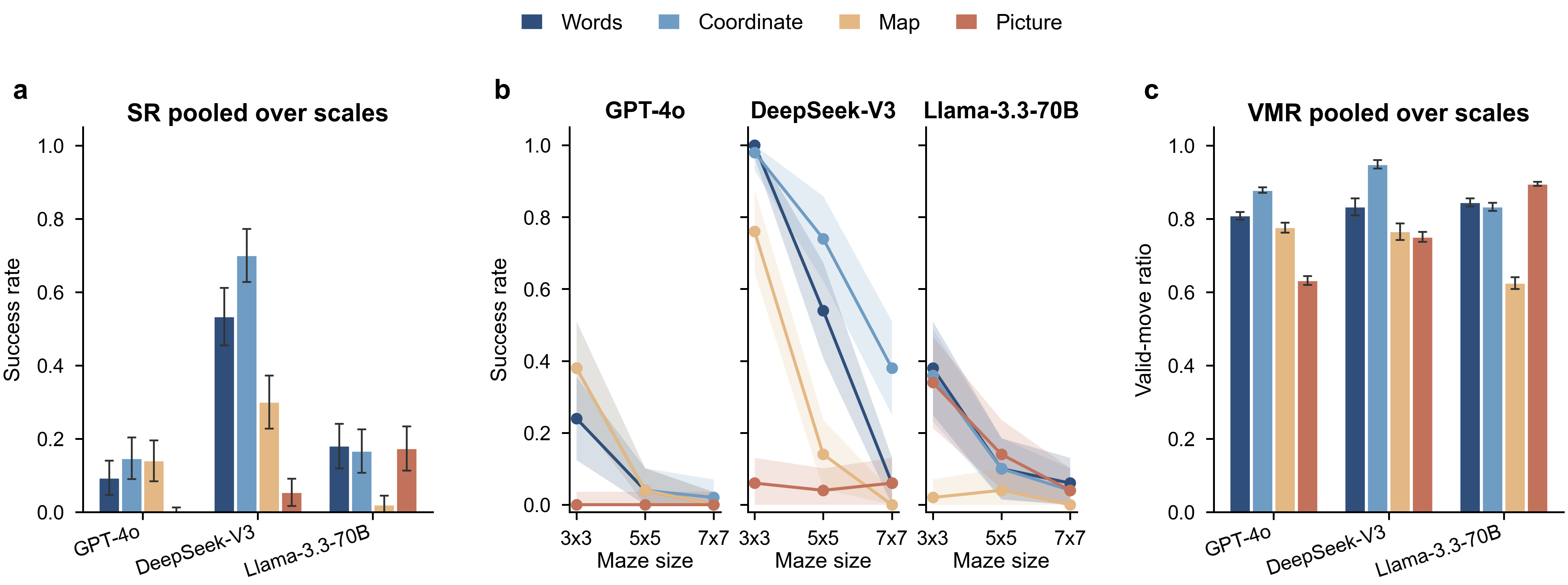}
    \caption{\textbf{Structured textual input outperforms visual and grid
    input across every model and size.} \textbf{(a)}~SR pooled over sizes
    $3,5,7$ by model and format ($n=150$ per bar; SR error bars are Wilson
    $95\%$ CIs, VMR bars $\pm 1$ SEM).
    \textbf{(b)}~SR vs.\ maze size, one panel per model. \textbf{(c)}~Valid-move
    ratio (VMR) pooled over sizes: high for every model--format pair,
    showing the SR collapse is a path-assembly failure, not a local-move
    failure.}
    \Description{Three-panel quantitative figure showing success rate by
    format and model, success rate versus maze size, and valid-move ratio
    by format and model.}
    \label{fig:rq1}
\end{figure*}

Module~1 addresses the input-acquisition stage (Figure~\ref{fig:overview},
module~\ding{172}). Figure~\ref{fig:rq1} and Table~\ref{tab:rq1} compare the
four input formats on small mazes. Structured \emph{Coordinate} text is the strongest format in
aggregate (mean SR $0.34$, vs.\ Words $0.27$, Map $0.15$, Picture $0.07$) and
ranks at or near the top for every model. The advantage is largest for the
strongest model, DeepSeek-V3, which reaches SR $0.70$ with Coordinate input
(Wilson $95\%$ CI $[62,77]$), more than ten times its $0.05$ ($[3,10]$) with
the rendered Image, a gap larger than sampling noise. At the per-model,
per-format level the smaller gaps are within noise (e.g., for GPT-4o,
Coordinate $15\%$ $[10,21]$ vs.\ Map $14\%$ $[9,21]$ overlap, a one-trial
difference at $n=150$). We therefore claim only the aggregate ordering
\mbox{Coordinate $\gtrsim$ Words $\gg$ Map $\gtrsim$ Picture} and DeepSeek-V3's
clean Coordinate-over-Image separation, not a strict per-model ranking. The
aggregate ordering holds at every size and is largest at $3{\times}3$ before all
formats decline toward zero by $7{\times}7$.

Valid-move ratio stays high
($\geq 0.62$, mostly $\geq 0.75$) for every model--format pair even where SR
is near zero. The models almost always emit grid-legal local moves, so the
collapse in SR is a failure of \emph{global path assembly}, not of local
move legality. This early dissociation (legal steps but illegal
routes) foreshadows the central finding of Module~2. We fix Coordinate as
the input format for Modules~2 and~3.

\begin{table}[t]
\caption{Module 1: one-shot SR (\%) by model and input format, pooled over
sizes $3,5,7$ ($n=150$ per cell). Best format per model in \textbf{bold}.}
\label{tab:rq1}
\centering
\small
\begin{tabular}{@{}lcccc@{}}
\toprule
Model & Words & Coordinate & Map & Picture \\
\midrule
GPT-4o        & 9  & \textbf{15} & 14 & 0 \\
DeepSeek-V3   & 53 & \textbf{70} & 30 & 5 \\
Llama-3.3-70B & \textbf{18} & 17 & 2 & 17 \\
\midrule
Mean          & 27 & \textbf{34} & 15 & 7 \\
\bottomrule
\end{tabular}
\end{table}

\subsection{Module 2: Where Do LLMs Fail Across Scales? (RQ2)}
\label{sec:res-rq2}

Module~2 addresses the multi-scale-representation stage
(Figure~\ref{fig:overview}, module~\ding{173}), probing each cognitive level
in isolation and inside full navigation.

\subsubsection{End-to-end navigation collapses early.}
\label{sec:res-rq2-overall}
Figure~\ref{fig:rq2main}a shows one-shot SR
versus maze size. Every model collapses to near zero by $10{\times}10$:
DeepSeek-V3, the strongest, falls from $0.96$ at $3{\times}3$ to
$0.06$ at $10{\times}10$ and $0$ thereafter. GPT-4o and Llama-3.3-70B
are already at or below $0.02$ by $7{\times}7$. Taken alone this only
restates that LLMs fail at larger mazes. The diagnostic question is whether
this reflects a collapse of local perception, of topology, of global
heading, or of their integration.


\paragraph{Size, not difficulty, is the dominant axis.}
Maze difficulty barely shifts where navigation breaks. Pooling over models,
the easy, medium, and hard tiers collapse at nearly the same rate once size
grows. At $5{\times}5$ all three sit at $30$--$32\%$ SR and at $7{\times}7$
at $10$--$14\%$, despite large differences in branching complexity,
dead-end density, and decoy length. Only at the smallest $3{\times}3$ does
easy ($61\%$) separate from medium and hard ($53\%$ each). Difficulty thus
sets the small-size ceiling but does not move the cliff; \emph{size}
does. We therefore run the full diagnostic sweep at medium difficulty
without loss of generality. The three models differ in level but not in
shape: DeepSeek-V3 sustains non-trivial navigation roughly one size step
further than GPT-4o and Llama-3.3-70B. Yet all three reach the same
near-zero floor by $10$--$15$.

\subsubsection{Isolated levels survive; coupling does not.}
\label{sec:res-rq2-isolation}
\begin{figure*}[t]
    \centering
    \includegraphics[width=\linewidth]{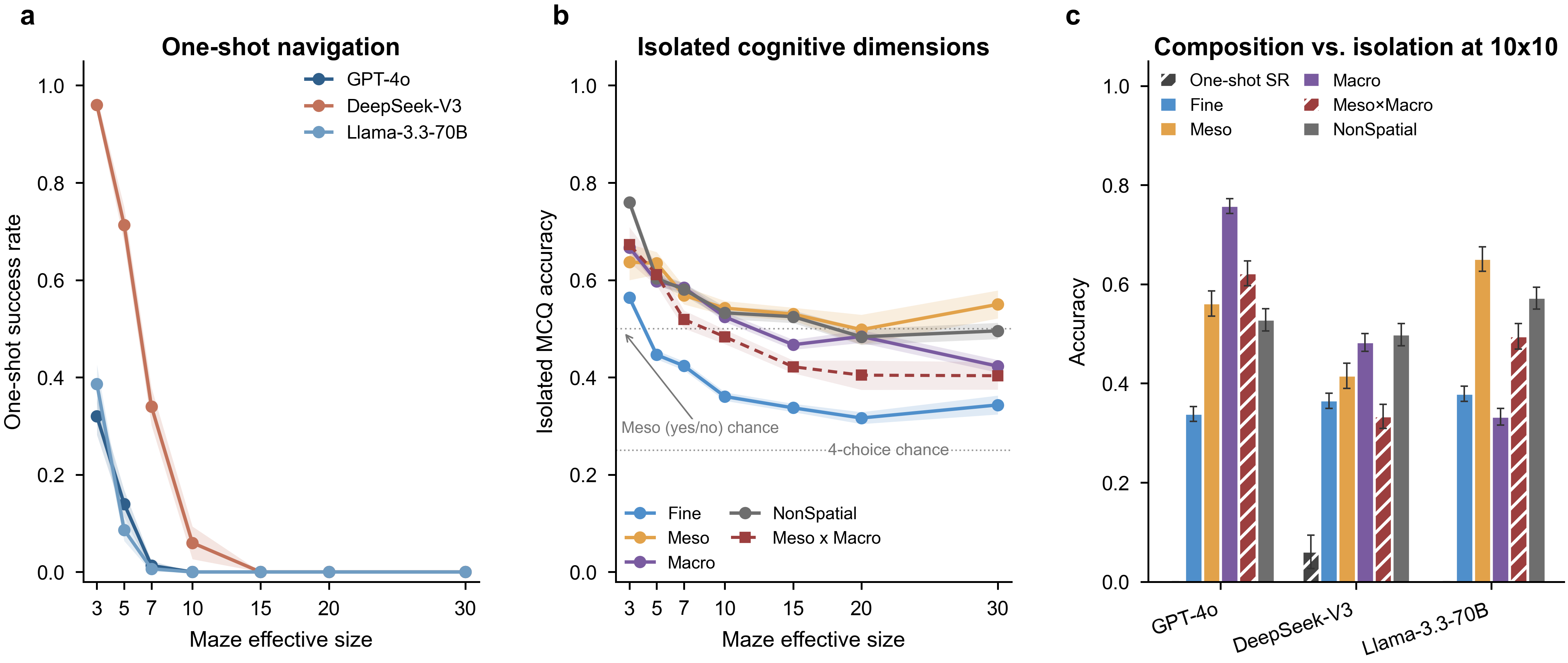}
    \caption{\textbf{Scaling failure is driven by aggregation, not local
    perception.} \textbf{(a)}~One-shot SR vs.\ size for three models; all
    collapse to $\approx 0$ by $10{\times}10$. \textbf{(b)}~Isolated probe
    accuracy (averaged over models): single-level Fine, Meso, and Macro
    probes persist, while the coupled Meso$\times$Macro probe (dashed)
    decays faster than its Meso component, comparably to Macro. \textbf{(c)}~Composition vs.\ isolation at $10{\times}10$:
    one-shot SR (hatched) has collapsed to $\approx 0$ while each isolated
    level remains at $0.3$--$0.75$. Bands: Wilson $95\%$ CI.}
    \Description{Three-panel figure: success rate versus size, isolated
    level accuracy versus size, and a grouped-bar comparison of one-shot
    success against isolated accuracy at the ten-by-ten size.}
    \label{fig:rq2main}
\end{figure*}

Figure~\ref{fig:rq2main}b--c answers the
diagnostic question. When each level is probed \emph{in isolation},
accuracy stays well above the navigation floor at \emph{every} size: Fine
(passable directions) holds $32$--$56\%$, Meso
(dead-end recognition) $50$--$64\%$, and Macro
(midcourse direction) $42$--$67\%$, never dropping to chance even
at $30{\times}30$. At $10{\times}10$, where one-shot SR is already
$0$--$6\%$, the isolated levels remain at roughly $30$--$75\%$. The
single most informative curve is the \emph{coupled} Meso$\times$Macro probe
(junction branch), which asks for the goal-reaching branch at a
junction and therefore requires topological reasoning \emph{and} allocentric
direction in one question. It decays faster than its Meso component and comparably to its Macro component, from $67\%$ at $3{\times}3$ to $\approx 40\%$ at $20{\times}20$--$30{\times}30$. The gap between
(i)~isolated competences that persist and (ii)~end-to-end navigation that
has already collapsed is direct evidence that the binding constraint is the
\emph{aggregation} of levels across a long sequential plan, not the
availability of any single level.

This dissociation is not an artifact of pooling over models: it holds for
each model individually (Table~\ref{tab:permodel}). At $10{\times}10$, where
one-shot SR has fallen to $0$--$6\%$, every model still answers the isolated
probes well above the $0.25$ chance level. GPT-4o is the clearest case: its
Macro heading probe stays at $76\%$ even as its end-to-end navigation reads
$0\%$. In an exploratory, small-sample test the dissociation also extends
beyond chat models: a reasoning model (o3-mini) collapses by $15{\times}15$
yet keeps its isolated competences intact (Appendix Fig.~\ref{fig:supp-o3mini}).

\paragraph{Compounding alone does not explain the collapse.}
One might object that the dissociation is mechanically trivial. If a model
clears each junction correctly only $\sim\!50\%$ of the time in isolation,
chaining many independent decisions drives end-to-end success toward zero on
its own, with no \emph{aggregation} deficit required. This null model is not a
straw man. Raising the isolated junction-branch accuracy to the number of
junctions a trajectory must clear reproduces the order of magnitude of the
collapse: $0.52^{4.5}\!\approx\!0.05$ versus $0.10$ observed at $7{\times}7$,
and $0.48^{5.5}\!\approx\!0.02$ matches the observed $0.02$ at $10{\times}10$.

What compounding cannot produce is a second, size-growing effect we measure
directly: the \emph{same} junction decision is made less reliably
\emph{inside} the loop than in isolation, and the gap widens with size.
Isolated versus in-trajectory junction accuracy (Meso JA) falls
$52\%\!\to\!43\%$ at $7{\times}7$, $48\%\!\to\!30\%$ at $10{\times}10$, and
$40\%\!\to\!12\%$ at $30{\times}30$, a monotonically growing gap of $8$, $18$,
and $28$ points. Pure compounding predicts \emph{no} such gap, since each
junction would be an independent draw at the isolated rate. The growing
in-loop degradation is therefore direct evidence of an aggregation cost beyond
multiplicative chaining: as the plan lengthens, the model makes more decisions
and makes each one less reliably.

\begin{table}[t]
\caption{Module 2: the navigation--isolation dissociation at $10{\times}10$,
shown per model. One-shot SR has collapsed while every isolated probe
remains well above chance ($0.25$). Values are \%; the Macro column pools the
full Macro level (initial and midcourse heading).}
\label{tab:permodel}
\centering
\small
\begin{tabular}{@{}lcccc@{}}
\toprule
Model & One-shot SR & Fine & Macro & Meso$\times$Macro \\
\midrule
GPT-4o        & 0 & 34 & 76 & 62 \\
DeepSeek-V3   & 6 & 36 & 48 & 33 \\
Llama-3.3-70B & 0 & 38 & 34 & 50 \\
\bottomrule
\end{tabular}
\end{table}

The NonSpatial controls confirm this is not merely long-context decay.
\emph{Row listing} (a pure reading task over the same maze text) is solved
at $92$--$100\%$ at \emph{every} size, so the models can read the
large-maze encoding faithfully. \emph{Cell count} nonetheless collapses
($97\% \to 20\%$), showing that some non-spatial aggregation over the whole
grid is also hard. We therefore rest the spatial argument on the
\emph{differential} decay of the coupled spatial probe relative to the
single-level probes (Figure~\ref{fig:ladder}), not on an absolute
spatial-vs-nonspatial contrast. A per-subtype breakdown is given in
Appendix Figure~\ref{fig:supp-subtypes}.

\subsubsection{The coupling ladder.}
\label{sec:res-rq2-ladder}
\begin{figure}[t]
    \centering
    \includegraphics[width=\columnwidth]{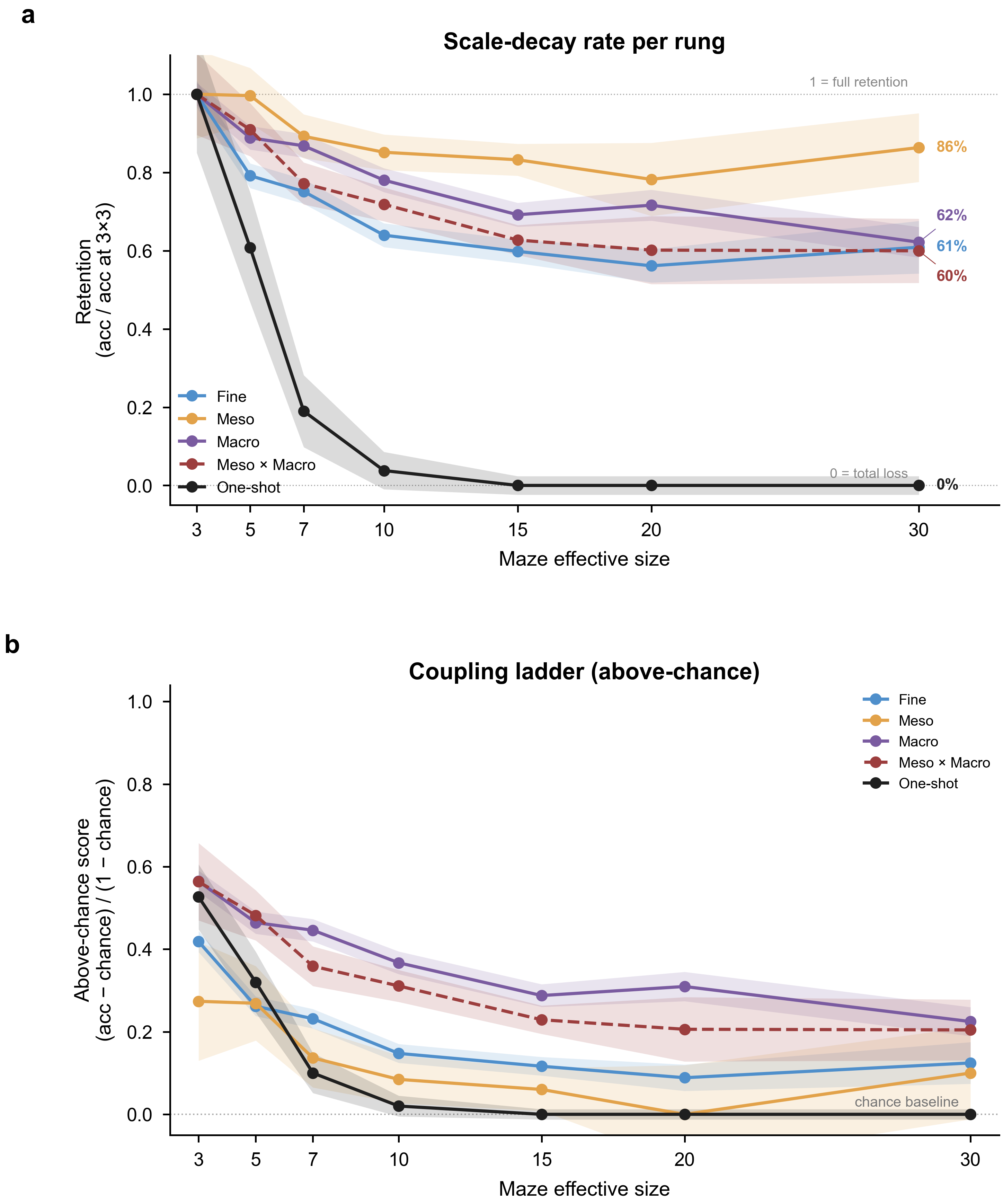}
    \caption{\textbf{The cost of each rung of coupling is dwarfed by the
    cost of moving from any single probe to full sequential navigation.}
    \textbf{(a)}~Accuracy normalized to each rung's own $3{\times}3$ value;
    the four single-question rungs retain $0.60$--$0.86$ at $30{\times}30$
    while one-shot navigation hits $0$ retention by $15{\times}15$.
    \textbf{(b)}~Above-chance score on a common scale, showing Macro and the
    coupled Meso$\times$Macro probe retain the largest margins at large
    sizes. Bands: Wilson $95\%$ CI.}
    \Description{Two stacked line charts showing normalized accuracy decay
    and above-chance score for five cognitive rungs versus maze size.}
    \label{fig:ladder}
\end{figure}

Figure~\ref{fig:ladder} arranges five ``rungs'' of increasing cognitive
coupling: Fine, Meso, Macro, coupled Meso$\times$Macro, and full one-shot
navigation. Each rung is normalized to its own $3{\times}3$ accuracy so
that slope visualizes size-decay. The four single-question rungs all retain
$60$--$86\%$ of their small-size accuracy at $30{\times}30$, whereas
one-shot navigation retains $0\%$ by $15{\times}15$. The qualitative jump is
not from one level to another but from \emph{any single multiple-choice
question} to \emph{multi-step sequential execution}: that transition is the
dominant scaling barrier. Panel~(b) re-plots all rungs as an above-chance
score $(\mathrm{acc}-\mathrm{chance})/(1-\mathrm{chance})$. This shows that
Meso's apparent durability is partly an artifact of its $0.50$ chance
baseline (only $0$--$27\%$ above chance), whereas the Macro and coupled
Meso$\times$Macro probes keep the largest genuine margins
($\sim\!20$--$30\%$ above chance) even at $30{\times}30$. The directional
signal is thus real and persistent, which makes its near absence from
first errors (below) all the more telling.

\subsubsection{Per-level degradation inside navigation.}
\label{sec:res-rq2-perlevel}
Beyond the binary collapse, per-trial metrics (Table~\ref{tab:rq2-perlevel},
Figure~\ref{fig:supp-perlevel}) show how each level erodes over the
\emph{whole} trajectory, and add three observations. First, Fine degradation
is dominated by teleports, not wall hits. The teleport rate climbs from $8\%$
at $3{\times}3$ to $38\%$ at $20{\times}20$, while the wall-collision rate
stays in single digits ($\leq\!8\%$). Positioning accuracy therefore falls
from $89\%$ to $\approx\!55\%$: the model loses track of its own position more
often than it walks into walls. Second, the model rarely recovers from a dead
end on its own. It enters dead ends at every size (DER peaks at $22\%$), but
the backtrack-success rate never exceeds $2\%$. Once inside a dead-end branch
it almost never retraces and takes another branch. Third, Macro stays the most
stable level inside navigation. The direction-drift rate holds at
$\leq\!12\%$ and the windowed progress rate declines only from $98\%$ to
$\approx\!74\%$, consistent with the isolated-probe results. Junction accuracy
(Meso JA), discussed above, falls from $74\%$ to $12\%$ and remains the
sharpest in-loop decline. The near-zero backtrack-success rate motivates
Module~3: handing dead-end retreat to a deterministic walker removes a failure
the LLM cannot fix on its own (Section~\ref{sec:res-rq3}).

\begin{table}[t]
\caption{Module 2: per-level diagnostic metrics (\%) measured \emph{inside}
one-shot navigation, by maze effective size (pooled over the three models,
medium difficulty). For PA, JA, and MPR higher is better; for WCR, TR, DER,
BSR, and DDR lower is better.}
\label{tab:rq2-perlevel}
\centering
\small
\begin{tabular}{@{}llccccccc@{}}
\toprule
Level & Metric & 3 & 5 & 7 & 10 & 15 & 20 & 30 \\
\midrule
\multirow{3}{*}{Fine}
 & WCR & 3 & 2 & 4 & 5 & 5 & 6 & 8 \\
 & TR  & 8 & 8 & 12 & 21 & 31 & 38 & 34 \\
 & PA  & 89 & 89 & 83 & 73 & 61 & 54 & 57 \\
\midrule
\multirow{3}{*}{Meso}
 & JA  & 74 & 57 & 43 & 30 & 24 & 13 & 12 \\
 & DER & 15 & 20 & 20 & 22 & 18 & 14 & 8 \\
 & BSR & 0 & 0 & 2 & 0 & 0 & 0 & 0 \\
\midrule
\multirow{2}{*}{Macro}
 & MPR & 98 & 96 & 89 & 85 & 78 & 72 & 74 \\
 & DDR & 0 & 4 & 8 & 8 & 12 & 7 & 7 \\
\bottomrule
\end{tabular}
\end{table}

\subsubsection{Where does the first error fall?}
\label{sec:res-rq2-firsterror}
\begin{figure}[t]
    \centering
    \includegraphics[width=\columnwidth]{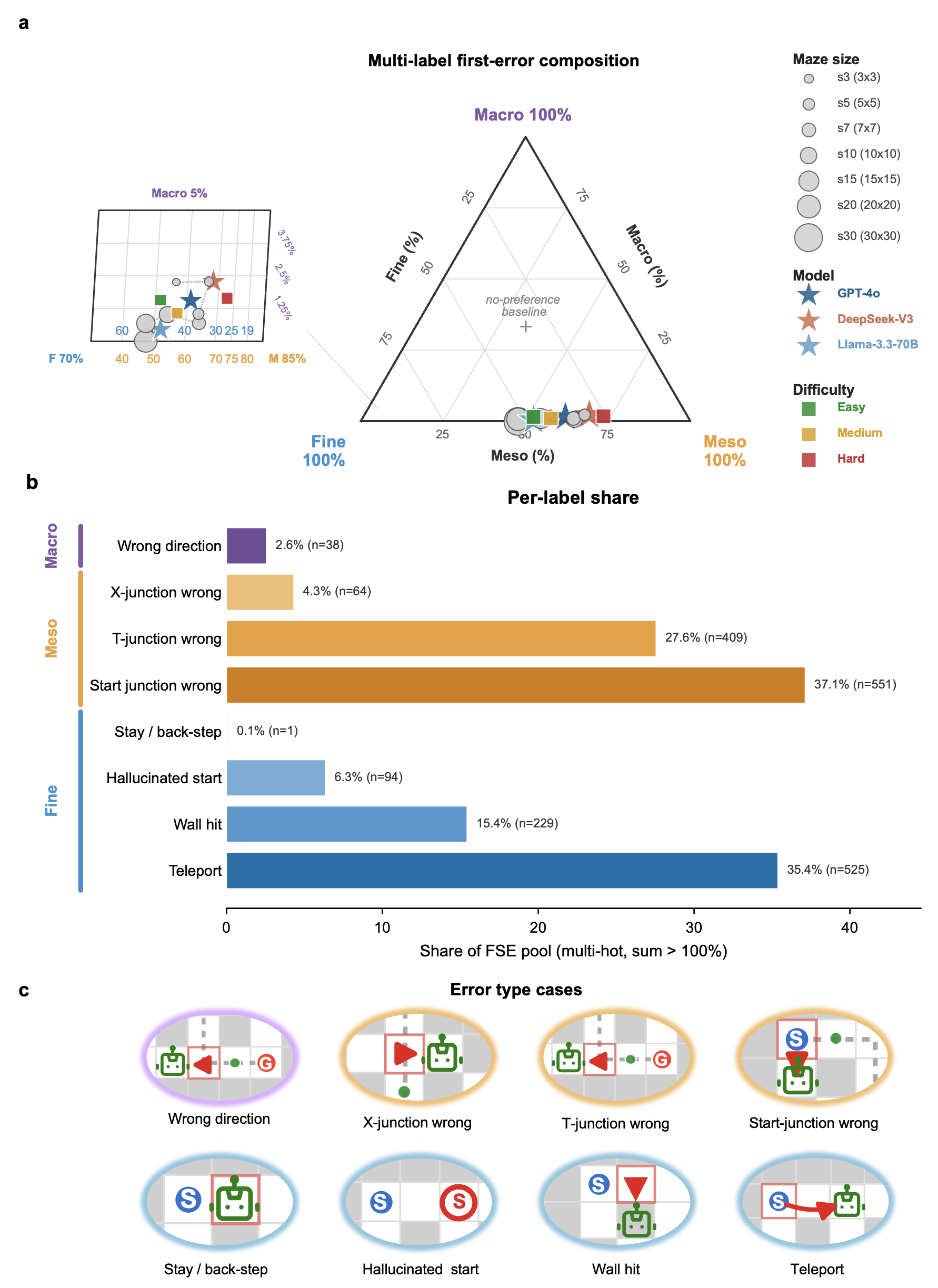}
    \caption{\textbf{First-step errors of failed one-shot navigation are
    dominated by Meso topological choices and Fine perception, with global
    direction rarely at fault} ($n=1{,}484$ failed trials, multi-hot
    8-label schema). \textbf{(a)}~Fine--Meso--Macro barycentric projection;
    the pool centroid lies at Fine $\approx 39\%$, Meso $\approx 59\%$,
    Macro $\approx 1\%$, and conditions cluster along the Fine$\leftrightarrow$Meso
    diagonal. \textbf{(b)}~Per-label share (bars do not sum to $100\%$ under
    multi-hot labelling). The icon strip illustrates each error type on a
    shared toy maze.}
    \Description{A ternary plot of first-error composition over Fine, Meso,
    and Macro levels, a per-label bar chart, and a strip of schematic
    icons illustrating each error type.}
    \label{fig:firsterror}
\end{figure}

The per-level rates above measure how much each level erodes over the
\emph{whole} trajectory. We now ask which level breaks \emph{first}. To
localize failure within a trajectory, we tag the first illegal or
sub-optimal move of every failed trial that produced at least one parseable
move under an eight-label, multi-hot schema grouped by level
(Figure~\ref{fig:firsterror}). Over the pool of $n=1{,}484$ failed trials,
the compositional centroid sits at \textbf{Fine $\approx 39\%$, Meso
$\approx 59\%$, Macro $\approx 1\%$}. First errors are dominated by Meso
topological choices and Fine perception, and global direction is almost
never the first thing to go wrong. By individual label, two Meso errors
(start-junction wrong $37.1\%$, T-junction wrong $27.6\%$) and two Fine
errors (teleport $35.4\%$, wall hit $15.4\%$) account for the bulk of
failures. Wall hits expose the gap between the two views. They are rare per
step (WCR $\leq\!8\%$, Table~\ref{tab:rq2-perlevel}) yet are the first break in
$15.4\%$ of failures, an infrequent move that is still decisive. The sole
Macro label (wrong direction) fires in only $2.6\%$ of trials. This
per-decision picture corroborates the aggregate isolation
result: the maintained global heading is repeatedly undone by mistakes at
branch points and in local execution as the plan lengthens.

Two caveats keep this from over-reading. First, ``Macro is rarely the
\emph{first} error'' is not ``heading never degrades''. The windowed Macro
progress rate still erodes within failed trials, but only gradually (cf.\ the
per-level MPR, Section~\ref{sec:res-rq2-perlevel}). A wall hit or wrong branch
usually precedes any measurable drift. The centroid thus localizes \emph{where trajectories first break}, which
structurally favors Fine/Meso over the slow-accruing Macro. Second, the
Meso/Fine dominance holds at every size but is not strictly size-invariant.
The Fine share rises with size (wall-hit share alone climbs $6\%\!\to\!37\%$,
$s3\!\to\!s30$) while Meso eases, so the balance tilts toward Fine in larger
mazes even as Macro stays negligible throughout. Models and difficulties
separate along the Fine$\leftrightarrow$Meso diagonal: Llama-3.3-70B and easy
mazes lean Fine (local execution), while DeepSeek-V3 and hard mazes lean Meso
(branch choices). The stronger model fails later, at genuinely topological
choices rather than basic perception.

\subsection{Module 3: Delegating Spatial Control (RQ3)}
\label{sec:res-rq3}

\begin{figure*}[t]
    \centering
    \includegraphics[width=\linewidth]{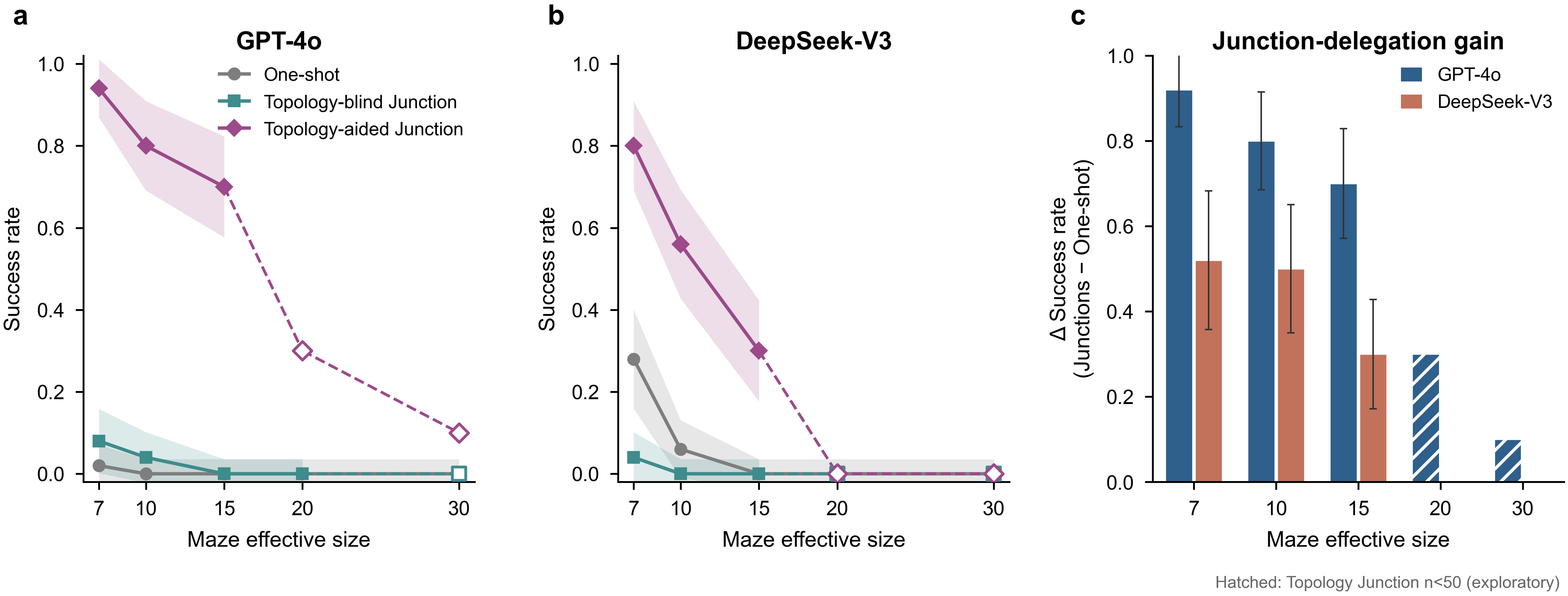}
    \caption{\textbf{Junction-level delegation lifts navigation but does not
    remove the scaling barrier.} \textbf{(a)}~GPT-4o and \textbf{(b)}~DeepSeek-V3
    SR under One-shot (grey), Topology-blind Junction (teal), and
    Topology-aided Junction (wine). \textbf{(c)}~Junction-delegation gain
    (Topology-aided minus One-shot) by model and size. Hollow markers /
    hatched bars mark exploratory cells with $n<50$. Bands: Wilson $95\%$ CI.}
    \Description{Two success-rate-versus-size line charts (one per model)
    comparing three delegation methods, plus a bar chart of the delegation
    gain by model and size.}
    \label{fig:rq3}
\end{figure*}

Module~3 addresses the hierarchical-route-planning stage
(Figure~\ref{fig:overview}, module~\ding{174}). Failure concentrates in
sustaining Meso/Fine decisions over a long plan, and the model almost never
backs out of a dead end unaided (BSR $\leq 2\%$; Table~\ref{tab:rq2-perlevel}). We therefore ask how much of that low-level
burden must be lifted from the LLM before navigation recovers.
Figure~\ref{fig:rq3} contrasts the three delegation regimes. Handing
corridor-walking and dead-end retreat to a deterministic algorithm, and
querying the model only at junctions \emph{with an explicit cell-type
prompt} (Topology-aided), transforms performance. GPT-4o rises from a
one-shot SR of $2/0/0\%$ at $7/10/15$ to $94/80/70\%$ (a lift of
$70$--$92$ points), and DeepSeek-V3 from $28/6/0\%$ to $80/56/30\%$, a lift
of $30$--$52$ points. Topology-blind, by contrast, barely moves SR: using
the \emph{same} walker, call sites, and navigation history but a
\emph{generic} prompt (no cell-type hint, no algorithmic dead-end retreat)
leaves SR within a few points of the one-shot baseline ($8/4/0\%$ for
GPT-4o), and below it for DeepSeek-V3 at $7{\times}7$. The lift is therefore
attributable to \emph{topological support} rather than to junction-level
decision granularity per se.

We are deliberate about what ``topological support'' resolves into. The aided
and blind regimes differ in two coupled ways (cell-type framing in the prompt
\emph{and} algorithmic dead-end retreat), so the design attributes the lift to
their combination, not to framing alone. The regimes are also cleanly matched
on realized LLM consultations only at $7{\times}7$ ($15.7$ vs.\ $17.4$ calls
per trial). At larger sizes the aided walker survives longer and issues more
calls ($114.6$ vs.\ $42.8$ at $15{\times}15$). We therefore treat $7{\times}7$
as the clean causal cell: there, equal call budgets with a $94\%$-vs-$8\%$ gap
show the gain is not bought by more queries. Yet recovery is bounded by size:
even Topology-aided SR declines by $20{\times}20$ and falls to near zero by
$30{\times}30$ (both exploratory $n<50$). Delegation buys roughly a doubling of
the navigable size; it does not remove the scaling wall.

\paragraph{What the walker telemetry reveals.}
Per-trial telemetry explains the gap (Table~\ref{tab:rq3-telem}). Under
Topology-aided delegation the algorithm physically retraces dead ends, so
the LLM faces only forward junction questions. GPT-4o reaches the goal in
$47/50$ mazes at $7{\times}7$ using only $\approx\!16$ junction decisions.
Its few failures are \emph{junction-budget exhaustion} on the largest
mazes (which demand over a hundred decisions) rather than getting trapped.
Under Topology-blind delegation the same model issues a comparable number of
calls but almost never finishes ($\leq\!8\%$). Lacking an explicit junction
cue and algorithmic retreat, it cannot stitch locally plausible moves into a
globally coherent route, and drifts away from the goal. Because the two
regimes are matched on architecture and call budget, the gap is attributable
to topological framing together with offloaded dead-end retreat.

\begin{table}[t]
\caption{Module 3 walker telemetry under Topology-aided delegation (GPT-4o,
medium): LLM junction decisions, dead-end branches retraced by the
algorithm, and total path steps. Failures at $7$--$15$ are budget
exhaustion, not entrapment.}
\label{tab:rq3-telem}
\centering
\small
\begin{tabular}{@{}lcccc@{}}
\toprule
Size & SR (\%) & Junction decisions & Dead-ends retraced & Path steps \\
\midrule
7  & 94 & 16  & 4  & 79 \\
10 & 80 & 49  & 10 & 234 \\
15 & 70 & 115 & 22 & 650 \\
\bottomrule
\end{tabular}
\end{table}

\section{Discussion and Conclusion}
\label{sec:discussion}

\paragraph{Format matters (RQ1).}
That a coordinate list beats a rendered image, and that local moves stay
legal even as whole-path success vanishes, suggests that, for current LLMs,
the useful spatial signal is symbolic and relational rather than pictorial.
For natural-language GIS and wayfinding interfaces, this argues for
surfacing spatial structure as explicit relations and coordinates rather
than relying on the model to parse rendered maps.

\paragraph{The level that does \emph{not} collapse first (RQ2).}
A natural hypothesis is that global heading (Macro) is the first casualty of
size. We find the opposite: Macro is the most durable level, both as an
isolated probe and in first-error analysis. Here ``durable'' means
\emph{slowest to fail}, not immune: within failed trials the windowed Macro
progress rate still erodes from $0.96$ to $\approx\!0.73$, just cumulatively
rather than as the opening error. What size destroys is the ability to
\emph{couple} a correct heading with correct topological choices step after
step. The Meso$\times$Macro probe decays faster than its Meso component
and comparably to its Macro component, and Meso junction errors dominate
failures. This is reminiscent of the human distinction between holding a sense
of direction and executing reliable route-level decisions
\cite{Wolbers2010, Montello1993}, and it reframes ``spatial reasoning
failure'' as a sequential-integration failure.

\paragraph{Why does aggregation fail?}
Our probes suggest a division of labor that current LLMs hold individually
but cannot sustain jointly. What degrades with size is the step-after-step
\emph{binding} of these signals: choosing the goal-consistent branch at each
junction while tracking which branches are already exhausted, much as a
bounded working memory erodes over a lengthening sequence. One-shot navigation
compounds this binding over dozens of dependent steps, which is why it
collapses an order of magnitude earlier than any isolated competence.

\paragraph{Designing hybrid spatial agents (RQ3).}
The delegation ladder gives a concrete recipe: an algorithm should own Fine
execution and dead-end retreat, the LLM should be consulted at topological
decision points, and the prompt must \emph{name} the topology explicitly.
This last step is the non-obvious one: the same architecture without
cell-type framing barely beats the baseline. The practical upshot for GIS
and routing agents is that LLMs are best used as junction-level
decision-makers inside an algorithmic skeleton, not as end-to-end path
generators.

\paragraph{Limitations.}
We flag five. \emph{(i)~Tree topology and external validity:} mazes are trees
(no cycles), so junction choices are unambiguous and we do not study
shortest-path optimization. This also bounds the hybrid-agent recommendation:
on cyclic, cost-weighted road networks a junction is a metric trade-off that
classical search (Dijkstra/A\textsuperscript{*}) solves optimally and more cheaply (a
$15{\times}15$ maze already needs $\sim\!115$ LLM calls per route). The
recipe therefore suits topology-dominated, semantic-disambiguation settings,
not metric routing, and we do not test cyclic graphs. \emph{(ii)~Confident failure:} from
Module~1, failures stay locally legal (VMR $\geq 0.62$ even where SR
$\approx 0$). A mis-navigating LLM emits plausible moves with no intrinsic
error signal, so safety-relevant deployments must verify each junction choice
against ground-truth topology. \emph{(iii)~Model and prompting scope:} three
chat-style models with fixed CoT, English only; a reasoning model (o3-mini)
is added only as a minimal supplement (Appendix Figure~\ref{fig:supp-o3mini}), and
alternative prompting and multilingual input remain out of scope. \emph{(iv)~Delegation
confound:} Module~3 is a delegation ladder, not a per-level oracle
ablation. Topology-aided and Topology-blind differ in two coupled ways
(cell-type framing \emph{and} algorithmic dead-end retreat), so the gap
attributes the lift to topological support without isolating framing from
retreat. \emph{(v)~Cognitive analogy:} Fine/Meso/Macro is a deliberate
metaphor, not a faithful instantiation. Our levels are representational
competences over one fully-observed symbolic maze (not Montello's
body-relative scale classes), we omit the landmark subsystem, and Macro is a
single goal \emph{bearing}, not survey knowledge. The largest Module~3 sizes
($20{\times}20$, $30{\times}30$) use $n<50$ and are exploratory.

\paragraph{Benchmark and release.}
We release the $1{,}050$ annotated mazes, the four input encoders, the
isolated-probe generators with answer keys, the junction-delegation harness,
and all evaluation and plotting code at the project page
(\url{https://yuhanjiang415.github.io/lost-in-aggregation/}). The annotation
already supports cyclic mazes and the harness accepts new per-level oracle
conditions, so future work can localize the binding bottleneck more finely
than the delegation ladder reported here.

\paragraph{Conclusion.}
We presented a multi-scale diagnostic benchmark that localizes \emph{where}
LLMs get lost in spatial navigation. The headline finding is that size does
not erode any single spatial competence so much as the ability to integrate
them across a long sequential plan. Junction-level delegation with explicit
topological framing recovers navigation at mid sizes but not at large ones,
pointing toward hybrid designs that pair LLM topological judgment with
algorithmic execution.

\begin{acks}

The authors used LLM-based writing tools (Anthropic Claude) to help edit and
polish the manuscript text. All experimental design, data collection,
analysis, figures, and scientific claims, as well as the substantive
writing, are the authors' own, and the authors take full responsibility for
the content of this work.
\end{acks}

\bibliographystyle{ACM-Reference-Format}
\bibliography{references}

\appendix

\renewcommand{\dbltopfraction}{0.9}
\renewcommand{\dblfloatpagefraction}{0.7}
\renewcommand{\topfraction}{0.9}
\renewcommand{\floatpagefraction}{0.7}
\renewcommand{\textfraction}{0.06}

\section{Appendix}

\begin{figure*}[!tb]
    \centering
    \includegraphics[width=\linewidth]{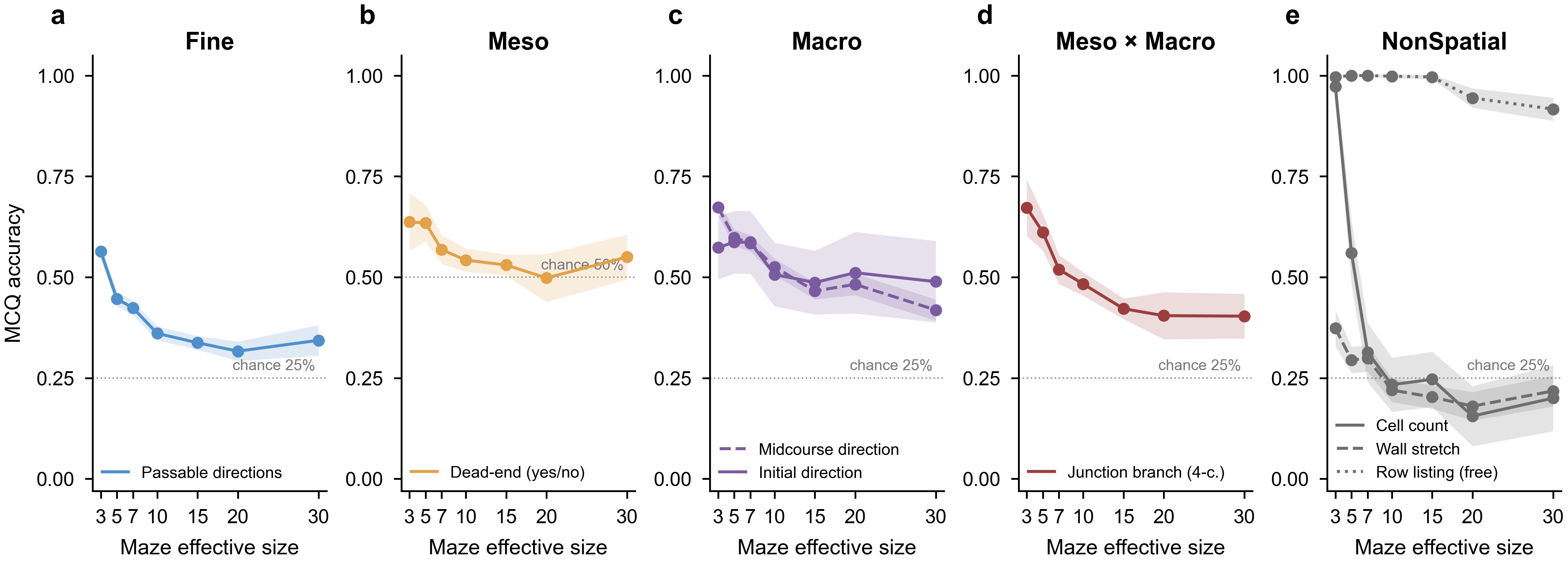}
    \caption{\textbf{Per-question-subtype accuracy across maze sizes}
    (isolated single-level probes, averaged over the three models). Facets:
    Fine, Meso, Macro, the coupled Meso$\times$Macro probe, and the NonSpatial
    control band; each plots accuracy versus maze effective size. The spatial
    subtypes stay above chance at every size, whereas the coupled
    Meso$\times$Macro probe decays fastest, matching the dissociation in the
    main text.}
    \Description{Five small-multiple line charts of accuracy versus maze size
    for each question subtype, grouped by cognitive level.}
    \label{fig:supp-subtypes}
\end{figure*}

\begin{figure*}[!tb]
    \centering
    \includegraphics[width=\linewidth]{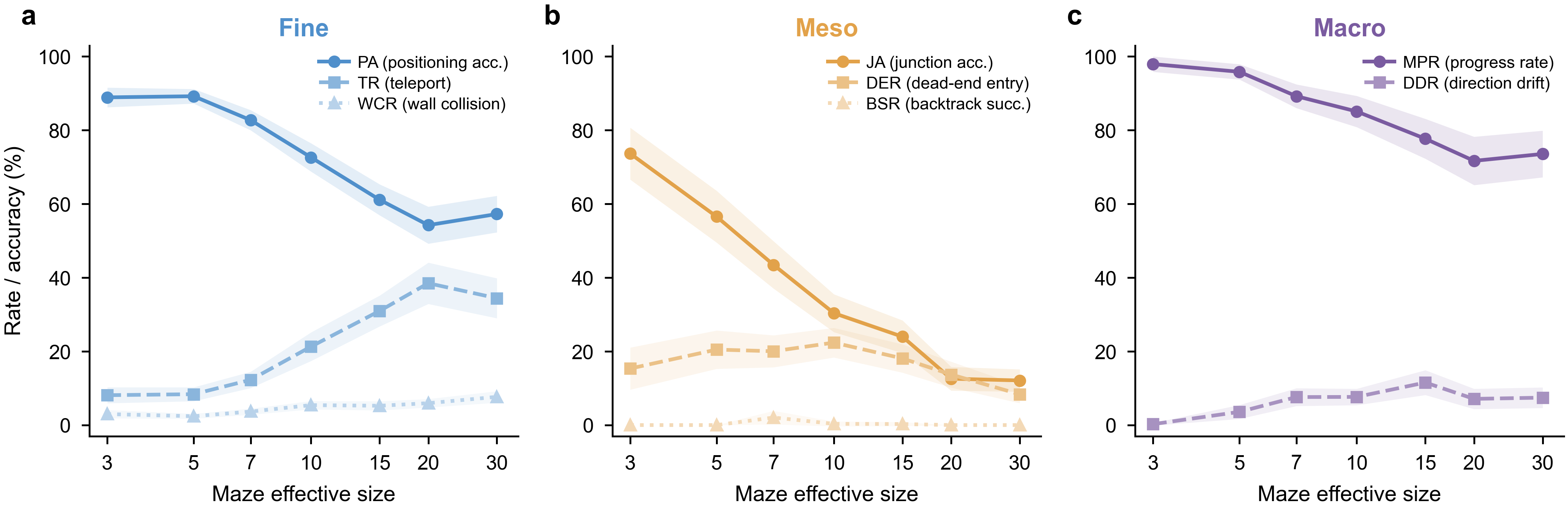}
    \caption{\textbf{Per-level diagnostic metrics measured \emph{inside} one-shot
    navigation, by maze effective size} (medium difficulty, pooled over the
    three models; the figure form of Table~\ref{tab:rq2-perlevel}).
    \textbf{(a)}~Fine: positioning accuracy (PA) falls as the teleport rate (TR)
    climbs, while wall collisions (WCR) stay rare. \textbf{(b)}~Meso: junction
    accuracy (JA) drops steeply; the model enters dead ends (DER) but its
    backtrack-success rate (BSR) stays near zero. \textbf{(c)}~Macro: the most
    stable level, with a slowly declining progress rate (MPR) and a low
    direction-drift rate (DDR). Bands: $95\%$ CI.}
    \Description{Three line-chart panels (Fine, Meso, Macro) of per-level
    navigation metrics versus maze size, showing teleport-driven Fine decay,
    a near-zero backtrack-success rate at the Meso level, and a stable Macro
    level.}
    \label{fig:supp-perlevel}
\end{figure*}

\begin{figure*}[!tb]
    \centering
    \includegraphics[width=0.76\linewidth]{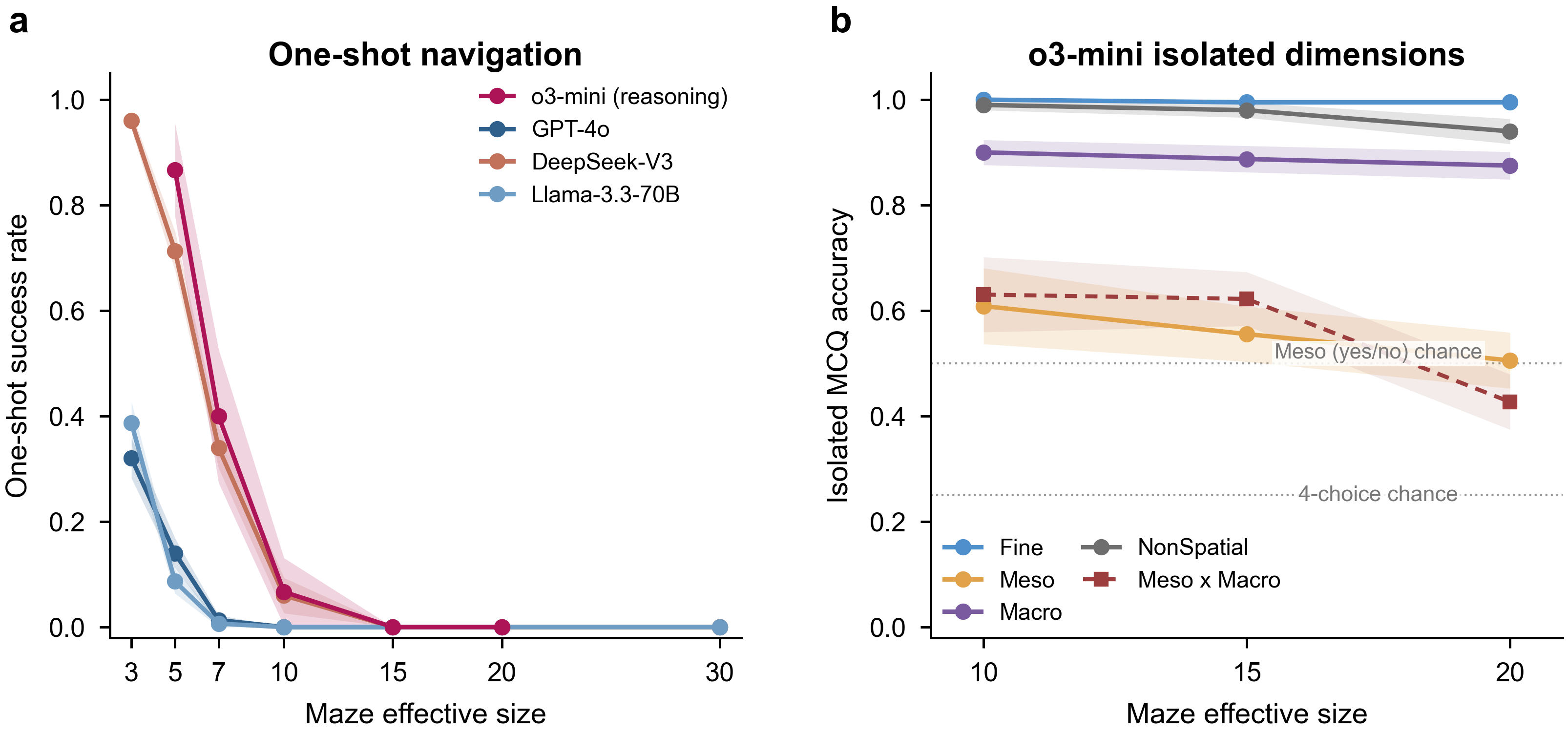}
    \caption{\textbf{A reasoning model (o3-mini) reproduces the main finding}
    (\texttt{reasoning\_effort} low, Coordinate input, medium difficulty;
    exploratory: $n$ small, single model). \textbf{(a)}~One-shot SR vs.\ size:
    o3-mini leads at small sizes (SR $0.87$ at $5{\times}5$, $0.40$ at
    $7{\times}7$; $n=15$) yet collapses on the same schedule as the chat models,
    to $0.07$ at $10{\times}10$ and $0$ by $15{\times}15$. \textbf{(b)}~Isolated
    single-level accuracy vs.\ size ($n\!\ge\!160$): Fine $\approx\!1$ and
    Macro $\approx\!0.88$ at $20{\times}20$ where one-shot SR is $0$, while the
    coupled Meso$\times$Macro probe decays fastest ($0.63\!\to\!0.43$, crossing
    chance). A reasoning model thus shifts the navigable frontier out by about
    one size step but does not remove the aggregation wall, confirming the
    paper's central claim that the bottleneck is the cross-scale aggregation of
    competences that remain individually intact. Bands: Wilson $95\%$ CI.}
    \Description{Two side-by-side panels: one-shot success rate versus maze size
    for o3-mini and three chat models, and o3-mini isolated multiple-choice
    accuracy versus size for Fine, Meso, Macro, NonSpatial, and the coupled
    Meso-by-Macro probe.}
    \label{fig:supp-o3mini}
\end{figure*}

\begin{figure*}[!tb]
    \centering
    \includegraphics[width=0.9\linewidth]{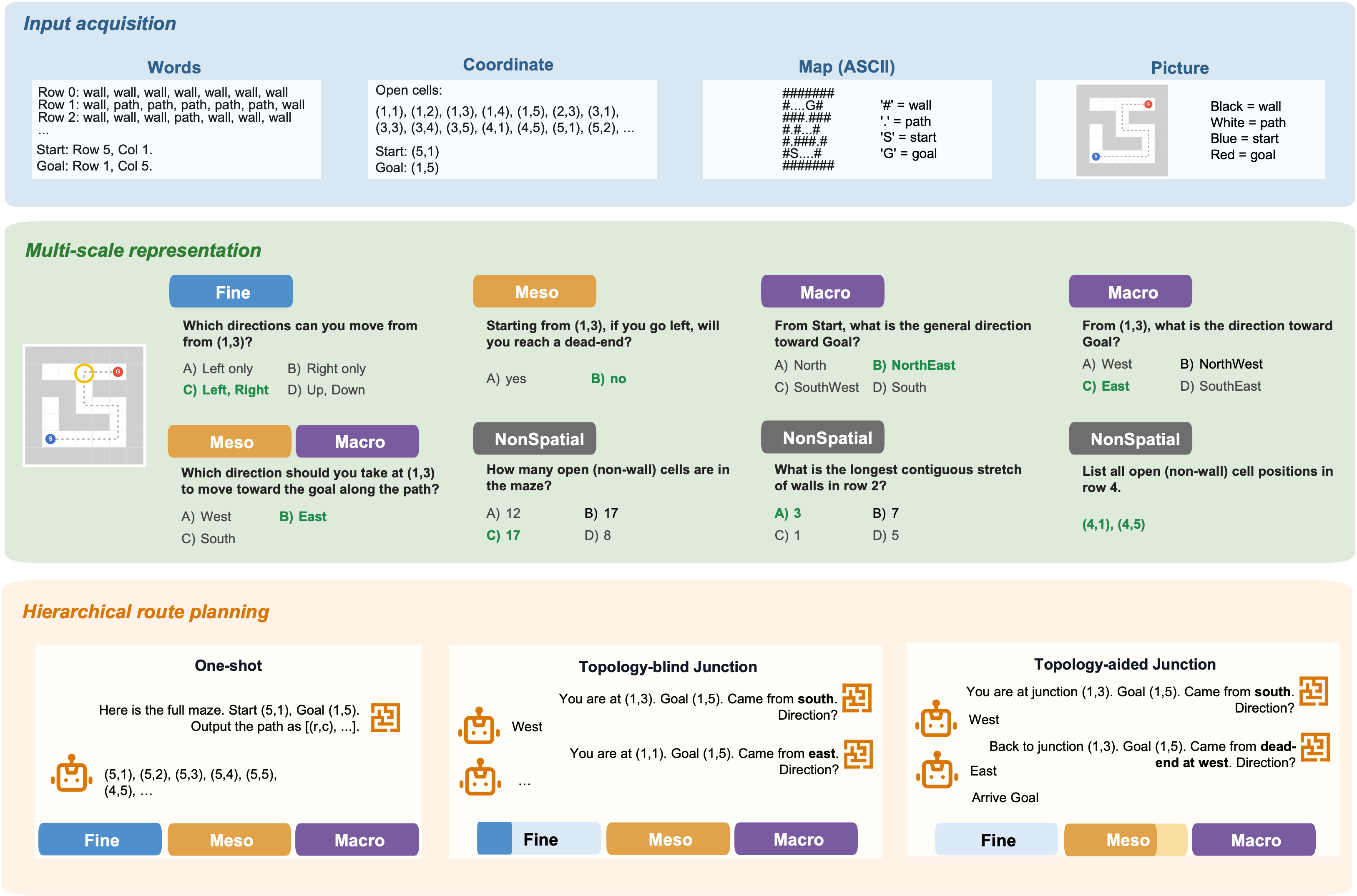}
    \caption{\textbf{Worked examples of the exact prompts and answer keys at
    every benchmark stage} for one small maze (start $(5,1)$, goal $(1,5)$):
    the four input formats (\emph{Input acquisition}), the isolated
    multiple-choice probes with the correct option marked (\emph{Multi-scale
    representation}), and the LLM exchanges for the three delegation regimes
    (\emph{Hierarchical route planning}).}
    \Description{A worked-example sheet showing concrete prompt text and answers
    for each input format, each isolated probe, and each delegation regime.}
    \label{fig:supp-framework}
\end{figure*}

This appendix collects four supplementary figures; full details are in each
caption. Figure~\ref{fig:supp-subtypes} breaks isolated-probe accuracy down by
question subtype, Figure~\ref{fig:supp-perlevel} reports the per-level
diagnostic metrics measured inside one-shot navigation,
Figure~\ref{fig:supp-o3mini} adds a reasoning-model (o3-mini) supplement, and
Figure~\ref{fig:supp-framework} shows worked examples of the prompts and answer
keys at every benchmark stage.

\end{document}